\documentclass{webofc}
\usepackage[varg]{txfonts}   
\begin{document}
\title{Interplay of baryonic chiral partners in fluctuations of net-baryon number density}
%
%

\newcommand{\dd}{\mathrm{d}}
\newcommand{\pp}{\partial}
\newcommand*\dif{\mathop{}\!\mathrm{d}}

\author{\firstname{Michał} \lastname{Marczenko}\inst{1}\fnsep\thanks{\email{michal.marczenko@uwr.edu.pl}} \and
        \firstname{Volker} \lastname{Koch}\inst{2,3}\and
        \firstname{Krzysztof} \lastname{Redlich}\inst{4,5} \and
        \firstname{Chihiro} \lastname{Sasaki}\inst{4,6}
}

\institute{Incubator of Scientific Excellence - Centre for Simulations of Superdense Fluids, University of Wroc\l{}aw, plac Maksa Borna 9, 50204 Wroc\l{}aw, Poland
\and
           Nuclear Science Division, Lawrence Berkeley National Laboratory, 1 Cyclotron Road, Berkeley, CA 94720, USA
\and
           ExtreMe Matter Institute EMMI, GSI Helmholtzzentrum für Schwerionenforschung, Planckstraße 1, 64291 Darmstadt, Germany
\and
            Institute of Theoretical Physics, University of Wroc\l{}aw, plac Maksa Borna 9, 50204 Wroc\l{}aw, Poland
\and
            Polish Academy of Sciences PAN, Podwale 75, 50449 Wroc\l{}aw, Poland
\and 
            International Institute for Sustainability with Knotted Chiral Meta Matter (WPI-SKCM$^2$), Hiroshima University, Higashi-Hiroshima, Hiroshima 739-8526, Japan
          }

\abstract{%
    In this contribution, we use the parity doublet model to investigate the fluctuations of the net-baryon number density. We discuss the systematics of the susceptibilities and their ratios for nucleons of positive and negative parity, as well as their correlator. We demonstrate that the fluctuations of positive-parity nucleon do not reflect the fluctuations of the total net-baryon number at the chiral phase transition.
}
\maketitle
\section{Introduction}
\label{sec:intro}

One of the prominent tasks within high-energy physics is to uncover the phase diagram of Quantum Chromodynamics (QCD), the theory of strong interactions. Throughout recent years experimental attempts were made to locate the critical point. Despite enormous experimental effort within the beam energy scan (BES) programs at the Relativistic Heavy Ion Collider (RHIC) at BNL~\cite{STAR2010} and the Super Proton Synchrotron (SPS) at CERN~\cite{Mackowiak-Pawlowska:2020glz}, this pressing issue remains unresolved (for a recent review see~\cite{Bzdak:2019pkr}).

The main theoretical and experimental probes of the critical point are fluctuations of conserved charges~\cite{Hatta:2003wn, Karsch:2010ck, Braun-Munzinger:2020jbk, Friman:2011pf}. Non-monotonic behavior for various fluctuation observables was observed in the Beam Energy Scan program which covered $\sqrt{s_{\rm NN}}=7.7-200~$GeV. In particular, indications of a non-monotonic behavior of the forth-to-second cumulant ratio of the net-proton multiplicity distributions in central Au+Au collisions were recently reported~\cite{STAR:2020tga}. However, experimental limitations force an assumption that the fluctuations of the net-proton number should reflect the overall fluctuations of the net-baryon number. The relation and differences between net-baryon and net-proton number fluctuations have not yet been explored, in particular, in theoretical models that consider dynamical chiral symmetry restorations in a strongly interacting medium.

One of the consequences of the restoration of chiral symmetry is the emergence of parity doubling around the chiral crossover. This has been recently observed in LQCD calculations in the spectrum of low-lying baryons around the chiral crossover~\cite {Aarts:2017rrl}. Such properties of the baryonic chiral partners can be described in the framework of the parity doublet model~\citep{Detar:1988kn, Jido:1999hd, Jido:2001nt}. The model has been applied to the vacuum phenomenology of QCD, hot and dense hadronic matter, as well as neutron stars~(see, e.g.,~\cite{Marczenko:2022hyt, Koch:2023oez} and references therein).

In this contribution, we apply the parity doublet model to calculate the susceptibilities of the net-baryon number distribution. Specifically, we focus on the fluctuations of individual parity channels and correlations among them. Their qualitative behavior is examined near the nuclear liquid-gas, as well as the chiral phase transitions.

\section{Fluctuations of baryonic chiral partners}
\label{sec:fluct}

To investigate the properties of strongly interacting matter, we use the parity doublet model~\cite{Detar:1988kn, Jido:1999hd, Jido:2001nt} for nucleons of positive ($N(939)$) and negative ($N^\star(1535)$~\cite{ParticleDataGroup:2022pth}) parity in the mean-field approximation~(see~\cite{Koch:2023oez} for detailed formulation of the model). 

The susceptibilities of the net-baryon number density can be calculated as follows~\cite{Koch:2023oez}. 
\begin{equation}\label{eq:x2_sum}
\chi_2^B = \chi_2^{++} + \chi_2^{--} + 2\chi_2^{+-} \rm,
\end{equation}
where $\chi_{++}~(\chi_{--})$ are the susceptibilities of the positive-(negative-) parity and $\chi_{+-}$ gives the correlations between them, i.e., correlations between vector densities. The susceptibilities are related to the event-by-event cumulants in the following way
\begin{equation}
\chi_2^{\alpha\beta} = \frac{1}{VT^3}\kappa_2^{\alpha\beta}\textrm, \;\;\;\; \chi_2^{B} = \frac{1}{VT^3}\kappa_2^{B} \textrm,
\end{equation}
for $\alpha,\beta=\pm$, where $\kappa_2^{\alpha\beta} = \langle \delta N_\alpha \delta N_\beta\rangle$, $\kappa_B = \langle \delta N_B \delta N_B\rangle$ are the cumulants, and $N_\alpha$'s' are the net number of positive/negative parity baryons ($N_B = N_+ + N_-$). Event-by-event cumulants and correlations are extensive quantities. They depend on the volume of the system and its fluctuations, which are unknown in heavy-ion collisions. The volume dependence, however, can be canceled out by taking the ratio of cumulants. Therefore, it is useful to define ratios of the cumulants of the baryon number. In the following, we focus on the ratios of the second and first-order cumulants of different parity distributions, which can be expressed through susceptibilities:
\begin{equation}
    R_{2,1}^{\alpha\beta} = \frac{\chi_2^{\alpha\beta}}{\sqrt{n_\alpha n_\beta}} \rm.
\end{equation}
where $n_\pm$ are the densities of the positive and negative parity states. The total net-baryon density is $n_B = n_+ + n_-$.
We note that in general the ratios, $R_{n,m}^{\alpha\beta}$, are not additive, e.g., $R_{2,1}^{++} + R_{2,1}^{--} + R_{2,1}^{+-} \neq R_{2,1}^B$.

\begin{figure*}[h]\label{fig:fig30}
\centering\includegraphics[width=0.46\linewidth]{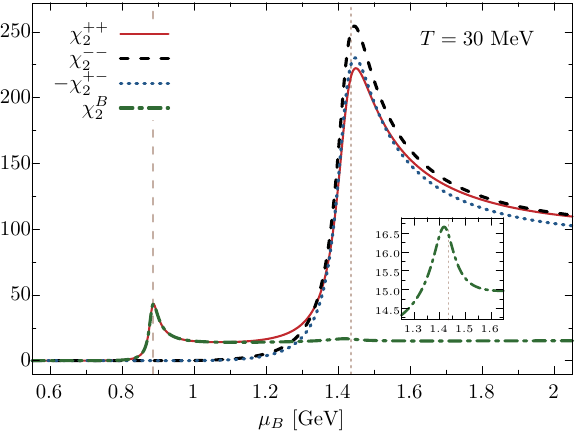}\;\;\;\;
\centering\includegraphics[width=0.46\linewidth]{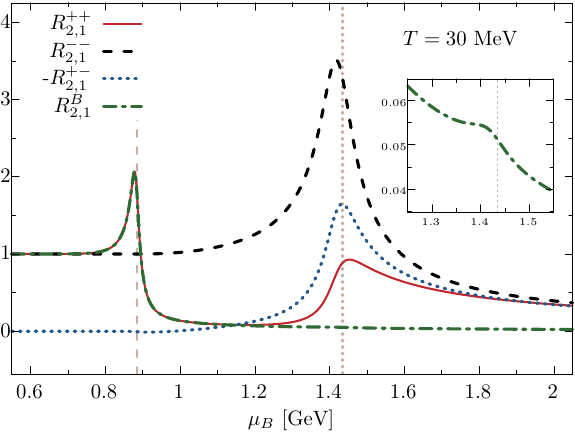}\;\;\;\;
\caption{Susceptibilities $\chi_2^{\alpha\beta}$ (left panel) and scaled variances $R_{2,1}^{\alpha\beta}$ (right panel) at $T=30$. Also shown are susceptibility, $\chi_2^B$, and scaled variance $R_{2,1}^B$ for the net-baryon number. We note that the correlator $\chi_2^{+-}$ and $R_{2,1}^{+-}$ are shown with the negative sign. The dashed and dotted vertical lines mark baryon chemical potentials for the liquid-gas and chiral crossover transitions, respectively. The inset figures show $\chi_2^B$ and $R_{2,1}^B$ in the vicinity of the chiral crossover transition.}
\end{figure*}

\begin{figure*}[h]\label{fig:fig100}
\centering\includegraphics[width=0.46\linewidth]{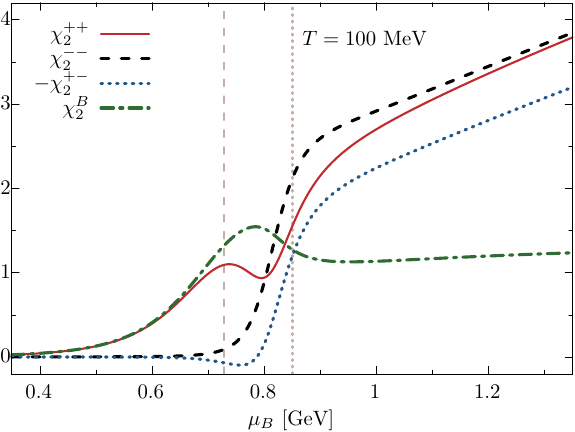}\;\;\;\;
\centering\includegraphics[width=0.46\linewidth]{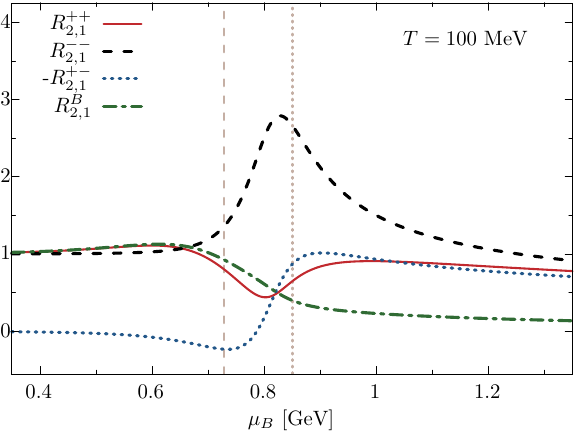}\;\;\;\;
\caption{The same as in Fig.~\ref{fig:fig30} but for $T=100~$MeV.}
\end{figure*}

In the left panel of Fig.~\ref{fig:fig30}, we show the susceptibilities $\chi_2^{\alpha\beta}$ for $T=30~$MeV. At $\mu_B < 1~$GeV, the net-baryon susceptibility develops a peak which is a remnant of the liquid-gas phase transition. The peak connected to the remnant of the chiral phase transition is developed around $\mu_B=1.4~$GeV. Due to thermal suppression of the negative-parity state, the net-nucleon susceptibility overlaps with $\chi_2^B$ at low $\mu_B$. At higher $\mu_B$ both $\chi_2^{++}$ and $\chi_2^{--}$ develop strong peaks connected to chiral symmetry restoration. At the same time, the correlator $\chi_2^{+-}$ becomes negative and develops a minimum of similar magnitude to the peaks in $\chi_2^{++}$ and $\chi_2^{--}$. Therefore, the negative correlation between baryonic chiral partners causes the suppression of the net-baryon number susceptibility around the chiral transition. In the right panel of Fig.~\ref{fig:fig30}, we show the scaled variances. Similarly to the susceptibilities, $R_{2,1}^B$ is suppressed around the chiral transition compared to individual ratios, owing to the negative correlator. Interestingly, the strongest peak at the chiral transition is developed by the $R_{2,1}^{--}$. At higher temperatures (see right panel of Fig.~\ref{fig:fig100}), the liquid-gas and chiral transition occur almost simultaneously. We observe that the structure of the individual susceptibilities and ratios does not necessarily reflect the critical structure for the total net-baryon number density. This qualitative difference is not only due to the presence of the negative-parity state but largely due to the non-trivial correlation between the chiral partners.

\section{Conclusions}
\label{sec:conc}

We have investigated the fluctuations of the net-baryon number density and discussed the qualitative role of the chiral criticality of hadronic matter. To this end, we used the hadronic parity doublet model. We have studied the susceptibilities of the positive- and negative-parity chiral partners, as well as their correlations. We find that the fluctuations in the vicinity of the liquid-gas phase transition are dominated by the positive-parity state. On the other hand, the vicinity of the chiral phase transition is characterized by large fluctuations of positive- and negative-parity states. We also find that, at the chiral phase boundary, the correlation between the chiral partners is non-trivial and negative. As a result, the fluctuations of the net-baryon number density are suppressed, compared to the positive-parity state fluctuations.

\section*{Acknowledgments}
This work is supported partly by the Polish National Science Centre (NCN) under OPUS Grant No.~2022/45/B/ST2/01527 (K.R. and C.S.) and the program Excellence Initiative–Research University of the University of Wroc\l{}aw of the Ministry of Education and Science (M.M.). The work of C.S. was supported in part by the World Premier International Research Center Initiative (WPI) through MEXT, Japan. K.R. also acknowledges the support of the Polish Ministry of Science and Higher Education. V.K. has been supported by the U.S. Department of Energy, Office of Science, Office of Nuclear Physics, under contract number DE-AC02-05CH11231, by the INT's U.S. Department of Energy grant No. DE-FG02-00ER41132, and by the ExtreMe Matter Institute EMMI at the GSI Helmholtzzentrum für Schwerionenforschung, Darmstadt, Germany.

\end{document}